# Room temperature fiber laser at 3.92 µm

FRÉDÉRIC MAES [1*], VINCENT FORTIN[1], SAMUEL POULAIN[2], MARCEL POULAIN[2], JEAN-YVES CARRÉE[2], MARTIN BERNIER[1] AND RÉAL VALLÉE[1]

[1]Center for Optics, Photonics and Lasers (COPL), Université Laval, Québec G1V 0A6, Canada
[2]Le Verre Fluoré, Campus KerLann, F-35170 Bruz, Brittany, France
*Corresponding author: frederic.maes.1@ulaval.ca

**Rare-earth-doped fiber lasers are promising contenders in the development of spectroscopy, free-space communications and countermeasure applications in the 3 – 5 µm spectral region. However, given the limited transparency of the commonly used fluorozirconate glass fiber, these systems have only achieved wavelength coverage up to 3.8 µm, hence fueling the development of more suitable fiber glass compositions. To this extent, we propose in this Letter a novel heavily holmium-doped fluoroindate fiber, providing extended transparency up to 5 µm, to demonstrate the longest wavelength room temperature fiber laser at 3.92 µm. Achieving ~ 200 mW of output power when cladding pumped by a commercial 888 nm laser diode, this demonstration paves the way for powerful mid-infrared fiber lasers emitting at and beyond 4 µm.** © 2018 Optical Society of America. Users may use, reuse, and build upon the article, or use the article for text or data mining, so long as such uses are for non-commercial purposes and appropriate attribution is maintained. All other rights are reserved.

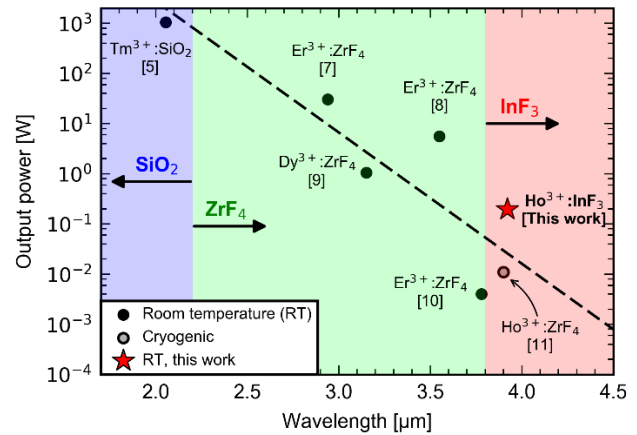

Fig. 1. Record continuous-wave output powers from room temperature RE-doped MIR FLs with respect to emitted wavelength.

Fiber lasers (FLs) find many applications in the medical, spectroscopy and manufacturing fields owing to their diffraction limited beam quality, as well as their rugged, maintenance-free and small footprint design [1]. However, extending the wavelength coverage of FLs in the mid-infrared (MIR) region, especially above 3 µm, while maintaining significant output power is an ongoing challenge. This spectral region has gathered much attention owing to the presence of fundamental molecular absorption bands enabling spectroscopy and remote sensing applications [2]. In addition, its overlap with the atmospheric transmission window at 3.9 µm is of particular interest for countermeasure and free-space communication applications [3,4].

Given the large number of optical transitions offered by rare-earth (RE) -doped glasses, RE-doped fibers have demonstrated significant wavelength coverage in the MIR, as shown in Fig. 1. At wavelengths around 2 µm, near-infrared diode pumped thulium (Tm³⁺)-doped silica FLs have reached the kilowatt output power level, an achievement made possible by the availability of high power fiber-based components as well as the high mechanical and thermal strength of silica fibers [5]. However, the limited transmission and high phonon energy (1100 cm⁻¹) of silica-based fibers render laser emission above 2.2 µm very unlikely [4]. Fluorozirconate (ZrF₄)-based fibers, on the other hand, possess a relatively low phonon energy of 574 cm⁻¹ that sets their infrared transmission edge around 4 µm [4].They have been the most successful in the demonstration of MIR laser emission above 2.4 µm, as seen in Fig. 1. Indeed, using core written fiber Bragg gratings (FBGs) [6] as well as single mode splices, near-infrared pumped erbium (Er³⁺)-doped monolithic all-fiber lasers have been demonstrated at both 2.94 µm and 3.55 µm, with output powers of 30 and 5.6 W respectively [7,8]. Additionally, 1.06 W at 3.15 µm has been demonstrated in a free-space in-band core pumped dysprosium (Dy³⁺)- doped ZrF₄ fiber laser [9] and 4 mW were achieved at 3.78 µm by stretching its limit the 3.5 µm transition in Er³⁺:ZrF₄ fibers [10]. Finally, the longest wavelength achieved from a fiber laser, to date, was reported two decades ago by Schneider *et al.* who demonstrated 11 mW of output power at 3.9 µm on the ⁵I₅→⁵I₆ transition of a holmium (Ho³⁺)-doped ZrF₄ fiber laser [11]. However, this demonstration had the major drawback of requiring both liquid nitrogen cooling as well as core pumping by a Ti:Sapphire laser to achieve threshold, hence marking the limit for RE-doped ZrF₄ fiber lasers in terms of MIR wavelength coverage. This limitation is mainly a consequence of the phonon-related properties of ZrF₄-based glasses that prevent laser emission to longer wavelengths for two main reasons. Firstly, because the emission of RE ions is quenched by multi-phonon (MP) decay, which increases exponentially with both temperature and emission wavelength [12]. Secondly, because the ZrF₄ glass' transparency rapidly falls off around 4 µm. Therefore, different RE-doped glass matrices have been sought in order to provide suitable optical properties for laser emission around and beyond 4 µm. Now, the recent availability of a new generation of low loss and heavily RE-doped InF₃ glass fibers with a transmission window extending up to 5 µm represents a crucial step towards a new generation of MIR fiber lasers.

Accordingly, we report here the longest wavelength room temperature diode-pumped fiber laser operating at 3.92 µm based on a novel holmium-doped fluoroindate glass (Ho³⁺:InF₃). Relying on a high Ho³⁺ concentration to enhance ion-pair energy transfer upconversion (ETU) processes and on excited state absorption (ESA) at the pump wavelength, the free-running cavity produces nearly 200 mW of output power with an optical-to-optical efficiency of 11%. This demonstration shows the benefits of using InF₃ fibers to unlock room temperature emission of long-wavelength transitions in RE ions and heralds a new generation of MIR FLs emitting near 3.9 µm and above.

The partial energy level diagram of the Ho³⁺:InF₃ system, along with relevant physical processes, is presented in Fig. 2. (a), where on the right hand side the lifetimes of the different energy levels are given. Ground state absorption (GSA) at 888 nm on the ⁵I₈ →⁵I₅ transition provides direct pumping of the upper laser level of the 3.9 µm transition [13]. Figure 2. (c) presents the cross-section of this transition which peaks around 888 nm at a value of 4.3 × 10⁻²⁶ m². Such pumping wavelength is readily available through high-power commercial multimode laser diodes and offers a simple and convenient approach to generate 3.9 µm emission in Ho³⁺:InF₃ fibers. Laser emission around 3.9 µm occurs between two excited levels of the Ho³⁺:InF₃ system on the ⁵I₅ → ⁵I₆ transition. The cross-section of this transition has been measured in bulk Ho³⁺:InF₃ as previously reported in [13] and is presented in Fig. 2. (b). One can see that it spans from 3840 to 4020 nm, overlapping the atmosphere's transmission window at 3.9 µm [3], and possesses a peak cross-section of 3.4 × 10⁻²⁵ m² around 3.92 µm. The lifetime, including radiative and non-radiative decay, of the upper laser level ⁵I₅ in Ho³⁺:InF₃ bulks has been measured to be 135 µs. For comparison, the measured lifetime of the same level in Ho³⁺:ZrF₄ bulks is 43 µs, a decrease mostly attributed to the higher phonon energy of ZrF₄ (574

cm⁻¹) compared to that of InF₃ (509 cm⁻¹) [14]. Nonetheless, the 3.9 µm transition in Ho³⁺:InF₃ glasses remains self-terminated since the lifetime of the lower level (⁵I₆) is 46 times longer than that of the upper level. However, recent spectroscopic studies have suggested that this limitation could be alleviated by using high Ho³⁺ concentrations to enhance ETU processes [15]. Moreover, excited state absorption (ESA) at 888 nm can also occur on the ⁵I₇ → ⁵F₅ transition [16], a process that has a two-fold effect on the 3.9 µm laser efficiency. Not only does it counteract ion bottlenecking in the long-lived ⁵I₆ level, it actually upconverts ions from the latter level to the ⁵F₅ level which then undergo MP decay to level ⁵I₅. Among the different energy transfers that have been reported in Ho³⁺:InF₃ [15], the one illustrated in Fig. 2. (a) (i.e. ⁵I₆, ⁵I₆→⁵I₅, ⁵I₈,⁵F₅) appears to be the most beneficial for laser emission at 3.9 µm. This ETU contributes to the population inversion of the 3.9 µm transition since it removes two ions from the lower energy level ⁵I₆ and recycles one of those ions back to the upper laser level ⁵I₅. A similar recycling mechanism is already exploited to increase the efficiency of the self-terminated ⁴I₁₁/₂ ⟶ ⁴I₁₃/₂ transition in highly-doped Er³⁺:ZrF₄ fiber lasers [17].

The schematic of the 3.92 µm room temperature fiber laser reported here is depicted in Fig. 3. The fiber cavity is made of a 23 cm long 10 mol. % Ho³⁺:InF₃ double-clad fiber developed by *Le Verre Fluoré*. A short length of fiber was selected to limit potential signal reabsorption from the lower level of the transition which can occur for insufficient pumping [13]. The measured fiber core's molar composition is 31 InF₃ - 30.5 BaF₂ - 19 ZnF₂ 9.5 SrF₂ - 10 HoF₃ while the cladding's molar composition is 41 InF₃ - 33 BaF₂ - 18 ZnF₂ – 8 SrF₂. The slightly multimode fiber core has a diameter of 16 µm and a numerical aperture (NA) of 0.2. The cladding has a circular diameter of 100 µm truncated by two parallel flats separated by 90 µm to enhance cladding pump absorption, and is coated with a low-index fluoroacrylate providing multimode guidance (NA > 0.4) at 890 nm. As seen in Fig. 4, the core attenuation of the drawn fiber is lying below 0.2 dB/m over the 3.4 – 4.0 µm spectral region, making the InF₃ fiber particularly suited for laser emission around 3.9 µm.

The Ho³⁺:InF₃ fiber cavity was bounded by two dichroic mirrors (DMs). The entrance DM1, providing 87% transmission at 888 nm and a broadband 99% reflectivity around 3.9 µm, was deposited on a quartz substrate while the output DM2 was fabricated on a ZnSe substrate with a reflectivity of 84% around 3.9 µm. The right-angled cleaved endfaces of the Ho³⁺:InF₃ fiber were secured in copper v-grooves with UV cured low-index polymer and were butted against the DMs using precision alignment stages. Optical pumping at 888 nm was provided by a multimode laser diode (LD, nLight element® e03) pigtailed to a 200/220 0.22 NA silica fiber. A set of lenses (L1, L2) in a 1:2 de-magnification configuration and gold mirrors (M1, M2) enabled the injection of the pump through the entrance DM1 into the cladding of the Ho³⁺:InF₃ fiber. Through a standard cutback measurement, the cladding absorption at 888 nm was measured to be 7.7 dB/m. An aluminum plate was used to passively cool the length of the fiber while fans provided forced convection to cool down the fiber tips protruding from the copper v-grooves.

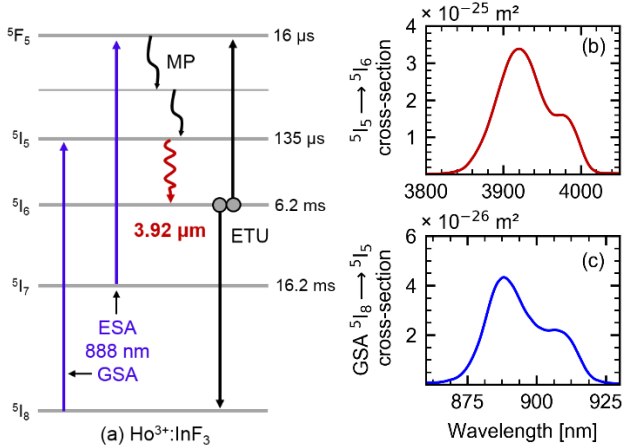

Fig.2. (a) Energy level diagram of the Ho³⁺:InF₃ system with relevant physical processes, (b) cross-section of the ⁵I₅ →⁵I₆ emission and (c) cross-section of the ⁵I₈ →⁵I₅ absorption reported in [13,15,16]. GSA, ground state absorption; ESA, excited state absorption; ETU, energy transfer upconversion.

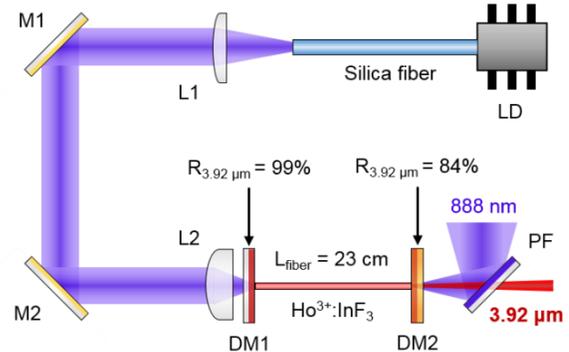

Fig. 3. Experimental setup of the room temperature fiber laser at 3.92 µm. LD, 888 nm multimode laser diode; L1 – L2, lenses in 1:2 de-magnification configuration; M1-M2, gold sputtered mirrors; DM1, home sputtered quartz dichroic mirror; DM2, home sputtered ZnSe dichroic mirror; PF, pump filter.

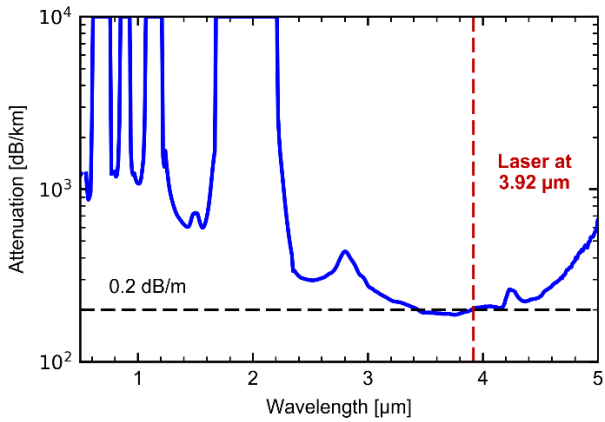

Fig. 4. Core attenuation of the Ho³⁺:InF₃ fiber from 0.5 to 5.0 μm measured by cutback.

The output power at 3.9 μm was measured with a low power thermopile detector (Gentec EO, model XLP12-3S-H2) along with a pump filter (PF) to reject residual pump power. The spectrum was analyzed by a mid-infrared optical spectrum analyzer (Yokogawa, model AQ6376 with extended wavelength coverage up to 5000 nm) at a spectral resolution of 0.2 nm.

The 3.92 μm output power as a function of the absorbed pump power at 888 nm is presented in Fig. 5. The laser threshold is located at 3.1 W and the optical-to-optical efficiency is around 11%, i.e. nearly half of the 22% Stokes efficiency. A record output power of 197 mW was achieved for an absorbed pump power of 4.9 W. Above this pump level, the cavity underwent failure at the butt-coupling between the fiber and the entrance DM1 due to an excess heat-load. Figure 6 displays the laser output spectrum for various output powers. As can be seen, the free-running cavity emits on four different Stark laser lines between 3917 and 3924 nm that are located near the peak of the emission cross-section measured in Ho³⁺:InF₃ bulks having an identical glass composition [13]. Broad wavelength scans did not reveal any spectral features near 2.1 and 2.9 μm, which could have been generated from transitions from lower lying energy levels. These transitions were possibly hindered by ESA and ETU processes illustrated in Fig. 2, to the benefit of the 3.9 μm transition.

When the pump diode was operated at low powers, the fluorescence emitted by the Ho³⁺:InF₃ fiber had a vivid red color. As the power of the pump was increased, the red fluorescence increased accordingly while additional visible components were seen to give more of a pink glow to the fiber. At the maximum absorbed pump power of 4.9 W, green visible fluorescence is clearly observed at the input of the Ho³⁺:InF₃ cavity. The evolution of the fiber color gives an insight into the kinetics of the local energy level populations. The constant red fluorescence emitted by the Ho³⁺ ions corresponds to spontaneous radiative decay to the ground state originating from the ⁵F₅ level, therefore indicating a population build-up in this level. This observation is in agreement with the energy level diagram depicted in Fig. 2 where ESA at 888 nm from the ⁵I₇ level and ETU from the ⁵I₆ level are seen to promote ions to the ⁵F₅ level. Moreover, the red fluorescence suggests that the contribution of ESA and ETU is increasing the efficiency of the 3.9 μm laser by enabling recycling of the pump excitation and limiting ion bottlenecking in lower lying energy levels (⁵I₇, ⁵I₆). As for the additional visible fluorescent components emitted by the fiber at higher pumping levels, spectroscopic investigations on Ho³⁺:ZrF₄ bulks suggest that they are the result of a second ESA at 888 nm originating from level ⁵I₅ [16]. In order to clearly assess the contribution of the different ESAs and ETUs on the 3.9 μm transition, additional spectroscopic investigations, supported by numerical modeling, need to be conducted.

Future investigations will also focus on the inscription of FBGs in the core of the InF₃ fiber by femtosecond laser [6]. Such implementation will enable power scaling of the system as it is compatible with more efficient passive cooling methods. Additionally, the combination of FBGs and low-loss splices will enable monolithic all-fiber cavity designs which have shown in the past unparalleled output power, efficiency and stability at 2.94 and 3.5 μm [7,8]. The power scaling of the Ho³⁺:InF₃ fiber laser will also be supported by the optimization of the InF₃ fiber. To this extent, much effort will be oriented towards improving its geometry, composition and drawing process to further reduce the background losses as well as increase the net gain at 3.9 μm. We believe that this research avenues will lead to the demonstration of watt-level monolithic Ho³⁺:InF₃ fiber laser cavities operating at 3.9 μm in the near future.

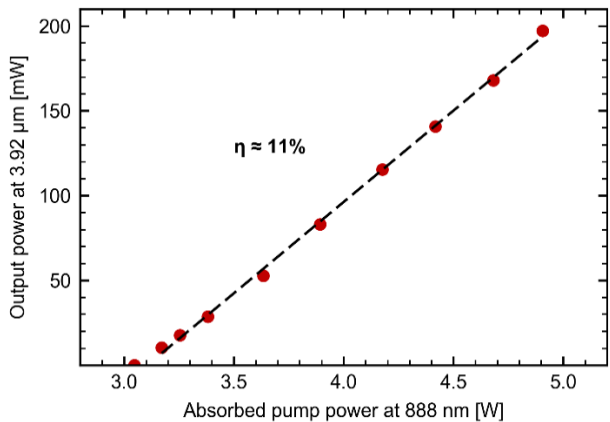

Fig. 5. Output power at 3.92 μm with respect to the absorbed pump power.

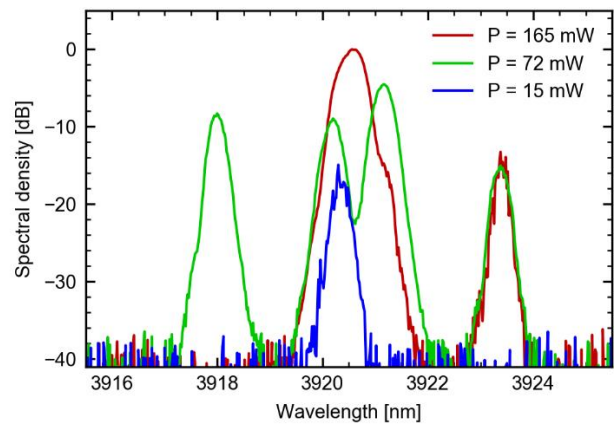

Figure 6: Spectrum of the Ho³⁺:InF₃ fiber laser for different output powers

Furthermore, the relative simplicity of the 3.9 μm fiber laser is likely to allow the development of gain-switched, Q-switched and mode-locked systems. Gain-switching has already been achieved in a Ho³⁺:InF₃ bulk with an identical composition and no foreseeable hurdle should limit the future demonstration of pulsed Ho³⁺:InF₃ fiber lasers [18]. In the case of mode-locking, non-linear polarization evolution has proven to be an efficient technique and has enabled the demonstration of 200 fs pulses with peak powers up to 37 kW around 2.9 μm in both Er³⁺ and Ho³⁺-doped cladding pumped ZrF₄ fiber cavities [20, 21]. This technique is a promising pathway to demonstrate mode-locking at 3.9 μm. Countermeasure applications, which seek modulated and highly directional MIR light sources, could largely benefit from the demonstration of pulsed fiber lasers in the atmospheric transmission window at 3.9 μm [3].

Finally, this demonstration has shown the benefits of the InF₃ glass matrix to unlock room temperature laser emission from a long wavelength MIR transition of the Ho³⁺ ion. Therefore, work will be devoted to studying and identifying other RE laser transitions that may, as of now, be accessible through the use of InF₃ fibers. For instance, praseodymium (Pr³⁺) offers an energy level arrangement where cascaded laser emission on two consecutive transitions around 4.7 μm ($^3H_6 \rightarrow ^3H_5 \rightarrow ^3H_4$) could be achieved through pumping at 1.95 μm with a Tm³⁺-doped fiber laser [21]. Alternatively, Dy³⁺-doped InF₃ has the potential to emit laser radiation around 4.3 μm ($^6H_{11/2} \rightarrow ^6H_{13/2}$) when pumped by a 1.7 μm Raman shifted erbium-doped fiber laser around 1.5 μm [22]. The demonstration of rugged fiber laser systems relying on these specific transitions will definitely contribute to the development of spectroscopy applications targeting molecular absorption bands between $4 - 5$ μm [2].

In summary, we have reported in this Letter the longest wavelength room temperature fiber laser. Based on a novel 10 mol. % Ho³⁺:InF₃ fiber, cladding pumped by a 888 nm commercial laser diode, the free-running cavity provides 197 mW of output power at 3.92 μm with an optical-to-optical efficiency of 11%, i.e. nearly half of the theoretical Stokes efficiency. This feat is enabled by the extended transparency (> 5 μm) and the lower phonon energy of the InF₃ fiber, as well as by excitation recycling processes enhanced by the high Ho³⁺ doping concentration. We believe that this demonstration will spark the development of a new generation of RE-doped InF₃ fiber laser systems operating at 3.9 μm, and beyond, which will address the unfulfilled needs of MIR applications in the $3 - 5$ μm spectral region.

**Funding.** Natural Sciences and Engineering Research Council of Canada (NSERC) (IRCPJ469414-13); Canada Foundation for Innovation (CFI) (5180); Fonds de Recherche du Québec-Nature et Technologies (FRQNT) (144616).

**Aknowledgements.** We thank the Yokogawa Test & Measurement Corporation for providing the OSA used for the spectral measurements, as well as Souleymane T. Bah and Marc D'Auteuil for their contribution in fabricating the dichroic mirrors.

## FULL REFERENCES